\documentclass[,trackchanges]{aastex701}

\usepackage{graphicx}
\usepackage[version=4]{mhchem}

\begin{document}

% \title{Nautilus Space Observatory: White Paper Template}

% \title{Nautilus: Fast time-resolved spectroscopic surveys of flares of GKM stars and their implications for planetary habitability.}

\title{Nautilus: Fast Time-Resolved Spectroscopy of GKM Stellar Flares and Their Implications for Planetary Habitability}

\author[0000-0001-5989-7594]{Chia--Lung Lin}
% \correspondingauthor{Chia--Lung Lin}
% \altaffiliation{Steward Observatory}
\affiliation{Steward Observatory, The University of Arizona, Tucson, AZ 85721, USA}
\email[show]{chialunglin@arizona.edu}

\author[orcid=0000-0000-0000-0001,sname='A.~D.~Fenstein']{Adina~D.~Feinstein}
\affiliation{Department of Physics and Astronomy, Michigan State University, East Lansing, MI 48824 USA}
\email{adina@msu.edu} 

\author[orcid=0000-0003-3305-6281]{Jeff Valenti}
\affiliation{Space Telescope Science Institute, 3700 Charles St., Baltimore, MD 21218}
\email{valenti@stsci.edu}

\author[0000-0002-2132-5264]{Mark S. Giampapa}
\affiliation{Steward Observatory, The University of Arizona, Tucson, AZ 85721, USA}
\affiliation{Lunar and Planetary Laboratory, The University of Arizona, Tucson, AZ 85721, USA}
\affiliation{National Solar Observatory, 950 N. Cherry Avenue, Tucson, AZ 85719, USA}
\email{giampapa@arizona.edu}

\author[orcid=0000-0003-3714-5855]{D\'aniel Apai}
\affiliation{Steward Observatory, The University of Arizona, 933 N. Cherry Avenue, Tucson, AZ 85721, USA}
\affiliation{Lunar and Planetary Laboratory, University of Arizona, 1629 E. University Boulevard, Tucson, AZ 85721, USA}
\affiliation{Alien Earths Team, NASA ICAR/NExSS, USA}
\email{apai@arizona.edu}

\author[orcid=0000-0003-2415-2191]{Julien de Wit}
\affiliation{Department of Earth, Atmospheric and Planetary Science, Massachusetts Institute of Technology, 77 Massachusetts Avenue, Cambridge, MA 02139, USA}
\email{jdewit@mit.edu}

\author[orcid=0009-0009-3020-3435]{Valeriy Vasilyev}
\affiliation{Max Planck Institute for Solar System Research, Justus-von-Liebig-Weg 3, 37077 G¨ottingen, Germany}
\email{vasilyev@mps.mpg.de}

\author[orcid=0000-0002-8842-5403]{Alexander Shapiro}
\affiliation{Max Planck Institute for Solar System Research, Justus-von-Liebig-Weg 3, 37077 G¨ottingen, Germany}
\affiliation{Institute of Physics, University of Graz, 8010 Graz, Austria}
\email{shapiroa@mps.mpg.de}

\author[orcid=0000-0002-8052-3893]{Prajwal Niraula}
\affiliation{Department of Earth, Atmospheric and Planetary Science, Massachusetts Institute of Technology, 77 Massachusetts Avenue, Cambridge, MA 02139, USA}
\affiliation{Kavli Institute for Astrophysics and Space Research, Massachusetts Institute of Technology, Cambridge, MA 02139, USA}
\email{pniraula@mit.edu}

\author[orcid=0000-0002-3627-1676]{Benjamin V. Rackham}
\affiliation{Department of Earth, Atmospheric and Planetary Science, Massachusetts Institute of Technology, 77 Massachusetts Avenue, Cambridge, MA 02139, USA}
\affiliation{Kavli Institute for Astrophysics and Space Research, Massachusetts Institute of Technology, Cambridge, MA 02139, USA}
\email[]{brackham@mit.edu}

\author[orcid=0000-0003-3989-5545]{Noah Tuchow}
\affiliation{Steward Observatory, The University of Arizona, Tucson, AZ 85721, USA}
\email{nwtuchow@arizona.edu}

\author[orcid=0000-0002-5322-2315]{Ana Glidden}
\affiliation{Department of Earth, Atmospheric and Planetary Science, Massachusetts Institute of Technology, 77 Massachusetts Avenue, Cambridge, MA 02139, USA}
\affiliation{Kavli Institute for Astrophysics and Space Research, Massachusetts Institute of Technology, Cambridge, MA 02139, USA}
\email{aglidden@mit.edu}

\author[orcid=0000-0002-6087-3271]{Nadiia Kostogryz}
\affiliation{Max Planck Institute for Solar System Research, Justus-von-Liebig-Weg 3, 37077 G¨ottingen, Germany}
\email{kostogryz@mps.mpg.de}

%\collaboration{all}{The Nautilus Space Observatory Collaboration}

\begin{abstract}

% The goal of the white papers is to identify compelling science questions that could be addressed by Nautilus, a constellation of large-diameter space telescopes. 

% Nautilus capabilities are not fixed yet: Starting from an initial trade space, we will refine the requirements through an iterative evaluation of the science questions, science requirements, instrument capabilities, and cost/schedule constraints. The white papers provide essential input for this process. 

% Please first focus on the science questions that may be suitable for Nautilus --- no need to demonstrate feasibility at this stage. After describing the science question, please complete the science requirements table to the extent it is possible now.

Low-mass GKM dwarfs are prime targets for finding habitable-zone Earth-sized planets, but their frequent flares, especially on M~dwarfs, can strongly affect planetary atmospheres through enhanced UV/XUV radiation and stellar proton events, which can drive complex photochemistry and accelerate atmospheric escape.
Current atmospheric and habitability models of planets around low-mass stars often rely on simplified flare inputs, such as fixed-temperature blackbodies or approximate optical-to-UV/XUV conversions. However, recent observations show that M~dwarfs' flare temperatures and spectral shapes can vary significantly with flare energy, phase, and stellar type, and that optical-based flare observations may underestimate the flare energy in the UV. Time-resolved spectroscopic flare observations of G- and K-dwarfs also remain rare compared to those of M dwarfs.
Here, we propose that the Nautilus Space Observatory concept can provide a unique opportunity to obtain fast-cadence, precisely flux-calibrated, moderate-resolution NUV-to-NIR spectroscopy of flares across a large sample of GKM dwarfs. These observations will measure the time-dependent flare energy budget from the near-UV/blue continuum to the optical and near-infrared continuum, while resolving key chromospheric lines that trace the underlying flare physics.
We aim to construct a statistical library of empirical flare spectral templates organized by flare and stellar properties, including flare energy, flare phase, and host-star spectral type. This library will provide a practical bridge between observed stellar flare properties and the radiation inputs required for planetary atmospheric evolution and habitability simulations.

\end{abstract}

% \keywords{}


%=========================
\section{Opening Statement}
This White Paper presents a potential science case for the Nautilus Space Observatory, a concept under development for a NASA Strategic Mission for the Astro 2030 Decadal Survey. Nautilus is a constellation of space telescopes and will provide a modular, scalable, sustainable, upgradable, expandable space observatory that can be deployed rapidly and then expanded progressively. The core concept for Nautilus is described in \cite{2019AJ....158...83A}. This White Paper is part of the first series of science white papers capturing ideas that emerged from the Nautilus Science Case workshop (held at MIT in May 2026).
% Recommended: 1--2 shorter paragraphs that describe overarching scientific goal and state of the art, leading to the problem statement and motivating the white paper.
\section{Scientific Context and Problem Statement} 
Low-mass GKM dwarfs are key targets for finding habitable-zone Earth-sized planets, but they are also among the most flare-active stars, especially M~dwarf stars \citep[e.g.,][]{2015ApJ...807...45D, 2024Sci...386.1301V, 2024AJ....168...60F, 2024AJ....168..234L}. 
Their flares can strongly affect planetary atmospheres through UV/XUV radiation, energetic particle impacts, ozone depletion, atmospheric escape, and climate changes \citep[][]{2010AsBio..10..751S, 2019AsBio..19...64T, 2025ApJ...985..100D, 2025AJ....170...40C}.

However, current habitability modeling often uses simplified flare inputs, such as fixed-temperature blackbodies of 9,000--10,000~K \citep[e.g.,][]{Kretzschmar2011} or approximate optical-to-UV/XUV conversions. 
These assumptions may be oversimplified because observed flare color temperatures can be cooler or hotter than the commonly assumed $\sim$9000--10,000~K value and can evolve throughout the flare phase \citep[e.g.,][]{2020ApJ...902..115H, 2022A&A...668A.111M, 2025AJ....170..297L}. 
% Moreover, optical-based flare models can underestimate the UV flare energy, with the canonical 9000~K blackbody model underpredicting GALEX NUV and FUV flare energies for M~dwarfs \citep[e.g.,][]{2023MNRAS.519.3564J}.
Moreover, optical-based flare models may not fully capture the high-energy flare emission that is most relevant for planetary atmospheres. 
For example, optical-based flare models can underestimate the UV flare energy, with the canonical 9,000~K blackbody model underpredicting GALEX NUV and FUV flare energies for M~dwarfs \citep[e.g.,][]{2023MNRAS.519.3564J}. Targeted multi-wavelength surveys of stellar flares have revealed a diversity of responses between FUV/NUV and optical emission \citep[e.g.,][]{2023MNRAS.519.3564J, 2024ApJ...971...24P, 2025AJ....169...27H, 2025ApJ...978...81K}.
More recently, archival HST/STIS time-tag Ly-$\alpha$ monitoring of TRAPPIST-1 has shown that frequent UV microflares can be revealed in time-tag UV spectroscopy, even when they are not clearly identified from optical observations alone, with much larger far-UV amplitudes than their optical counterparts \citep[e.g.,][]{2026ApJ...998L..31B}. 
Together, these results suggest that optical flare statistics alone are insufficient for characterizing the UV flare forcing experienced by close-in exoplanets .

In addition, time-resolved spectroscopic flare observations of G- and K-dwarfs remain very rare compared to those of M~dwarfs, with only a small number of active G/K stars, such as EK~Dra, LQ~Hya, and HD~189733, having been studied in this way \citep[e.g.,][]{1999MNRAS.305...45M, 2017A&A...607A..66K, 2022NatAs...6..241N, 2022ApJ...926L...5N}.
This motivates the need for a large, time-resolved spectral sample that can empirically characterize the evolution of flare temperatures and spectral shapes across low-mass stars on second-to-minute timescales.

The Nautilus Space Observatory concept---a very-large-aperture, ultralight space telescope for exoplanet exploration, time-domain astrophysics, and faint objects \citep{2019AJ....158...83A}---could provide a unique combination of wide-field monitoring capability, sensitivity, and fast-cadence NUV-to-NIR spectroscopy for a large sample of low-mass stars. These observations would enable more realistic flare inputs for exoplanet atmospheric escape, photochemistry, and habitability models.

% \textbf{Problem Statement: 1--2 sentences that identify the problem that requires new type of data.}

% We require the new type of data from Nautilus, the fast time-resolved, moderate resolution, and well-flux-calibrated spectroscopic of low-mass stars' flare from NUV-to-NIR, to address the following key questions:
% \begin{itemize}
% \item How is flare propagation through different layers of the stellar atmosphere reflected in chromospheric heating features, such as the Balmer jump, and the optical thermal properties of flares?

% \item Given these observations, how can we derive more realistic flare inputs for planetary atmospheric modeling and assess their implications for habitability?
% \end{itemize}

%=========================
\section{Science Objectives} 

% Please describe briefly what are the objectives of the program. This section may consist of a few bullet points with clearly stated objectives, followed by a one or two paragraphs of justification. 

We require fast time-resolved, moderate resolution, and flux-calibrated spectroscopy of stellar flares on low-mass stars from NUV-to-NIR. With these specifications, we will be able to address the following key questions:
\begin{itemize}
% \item How is flare propagation through different layers of the stellar atmosphere reflected in chromospheric heating features, such as the Balmer jump, and the optical thermal properties of flares and the properties of stars?
\item How is flare-energy deposition and propagation through different layers of the stellar atmosphere reflected in chromospheric heating diagnostics, such as the Balmer jump, and in the optical/near-infrared thermal properties of flares? 

Multi-line, high-cadence spectroscopy can probe different atmospheric depths, temperatures, and optical-depth regimes. 
By measuring the relative amplitudes, time lags, and line-profile evolution of photospheric, chromospheric, and upper-atmosphere tracers, Nautilus can constrain how flare energy is deposited and transported through stellar atmospheres as a function of flare energy, flare phase, and stellar properties across the GKM sequence.

% \item How are coronal mass ejections (CMEs) associated with flares of different energy levels across the GKM sequence? 
% While flares provide radiative energy input to planetary atmospheres, CMEs can drive particle impacts, atmospheric escape, and magnetospheric compression. 
% By combining high-cadence and high-resolution spectroscopic flare diagnostics with searches for blue-shifted line asymmetries, transient absorption/emission components, and velocity evolution in chromospheric tracers, Nautilus can test whether CME occurrence, velocity, and mass-loss signatures scale with flare energy, flare phase, and stellar properties. 
% This will help establish whether the solar flare--CME relation can be extended to low-mass stars, or whether magnetic confinement suppresses detectable CMEs in active M dwarfs.

\item How do flares impact atmospheric evolution and habitability of short-period exoplanets orbiting GKM stars?
% \item How can multi-line, high-cadence spectroscopy be used to map the atmospheric-layer-dependent response of stellar flares across the GKM sequence? Different diagnostics probe different depths, temperatures, and optical-depth regimes in the stellar atmosphere. By measuring the relative amplitudes, time lags, and line-profile evolution of photospheric, chromospheric, and upper-atmosphere tracers, Nautilus can constrain how flare energy is deposited and transported through stellar atmospheres as a function of flare energy, flare phase, and stellar spectral type.
\end{itemize}

\begin{figure*}[ht!]
\centering
\includegraphics[width=\textwidth, height=0.5\textheight, keepaspectratio]{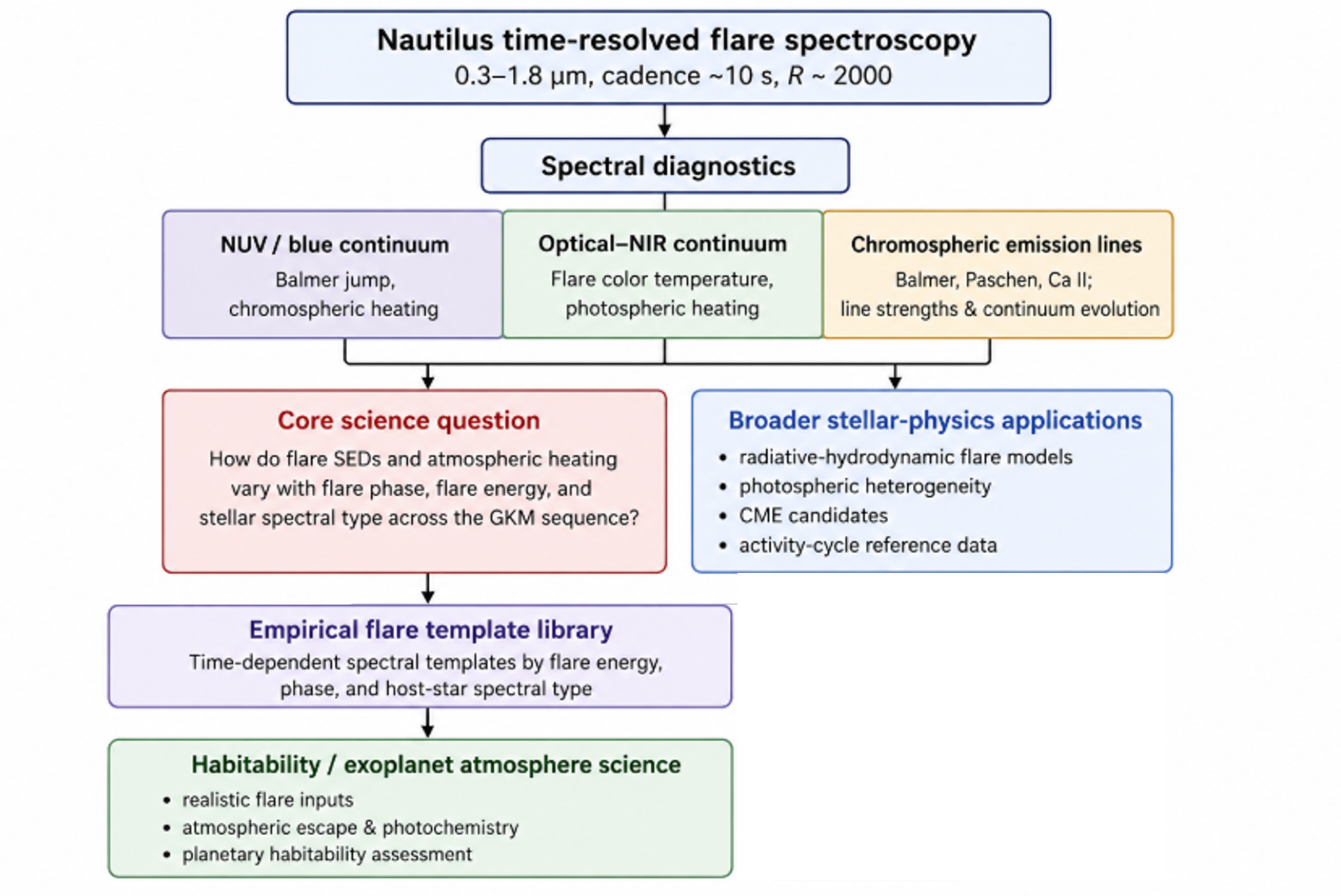}
% \caption{Example figure.}
% \label{fig:general}
\caption{Flowchart summarizing the proposed Nautilus time-resolved flare spectroscopy program.
Fast NUV-to-NIR spectroscopy will connect continuum and line diagnostics to the core science goal of measuring how flare spectral energy distributions and atmospheric heating vary with flare phase, flare energy, and stellar spectral type across the GKM sequence.
The resulting empirical flare-template library will provide time-dependent flare inputs for exoplanet-atmosphere modeling and will also support broader stellar-physics applications, including flare-heating model tests, stellar photospheric heterogeneity studies, CME-candidate searches, and activity-cycle reference measurements.}
\label{fig:science_flowchart}
\end{figure*}

A summarized flowchart illustrating the connection between the proposed Nautilus observations, the key spectral diagnostics, the core science question, and the main science applications is shown in Figure~\ref{fig:science_flowchart}. 
In the following sections, we describe in more detail the required data, proposed observing strategy, and analysis methods for connecting these flare observations to both stellar-flare physics and planetary-atmosphere modeling.

%=========================

%=========================
\section{Data Requirements}

To address these science objectives, we need fast time-resolved spectroscopic observations of flares from low-mass stars. 
The required wavelength coverage is from $0.3~\mu{\rm m}$ to $1.8~\mu{\rm m}$, covering the NUV to the NIR. 
The NUV and blue optical coverage are needed to measure chromospheric heating features such as the Balmer jump, while the optical and NIR coverage are needed to constrain the thermal spectral evolution of flares \citep[e.g.,][]{2013ApJS..207...15K, 2025ApJ...994L..31H}.  The ability to obtain time-resolved spectroscopy across this broad wavelength range is a unique capability of Nautilus that is unrivaled at any single facility.

% Julien:
In addition to measuring the broadband continuum evolution, the wavelength coverage should include multiple activity-sensitive diagnostics that probe different atmospheric layers. 
These include near-UV/blue continuum features and the Balmer jump, chromospheric lines such as the Balmer series, Ca~\textrm{II}~H~\&~K, H$\alpha$, and the Ca~\textrm{II} infrared triplet, and, where available through complementary UV observations or future extensions, upper-atmosphere diagnostics such as Ly$\alpha$. 
% A moderate spectral resolving power of $R\sim2000$ is sufficient for these science goals of this program. 
% Simultaneous or near-simultaneous monitoring of these tracers would allow us to connect photospheric/continuum heating, chromospheric response, and upper-atmosphere flare signatures within a single time-dependent framework.

% The targeted cadence, or sampling speed, is around 10 seconds. 
% This cadence is needed to capture the rapid spectral evolution during the flare phase. 
% In terms of data quality, we aim for a signal-to-noise ratio of ${\rm SNR}=10$ per spectral pixel at $0.4~\mu{\rm m}$ per exposure/sample.

A moderate spectral resolving power of $R\sim2000$ is sufficient for measuring these activity-sensitive diagnostics and for addressing the core science goals of this program.
This resolution is comparable to existing moderate-resolution spectroscopic surveys that have successfully measured flare and chromospheric activity diagnostics in low-mass stars, including SDSS time-resolved spectra of M-dwarf flares and LAMOST low-resolution measurements of Ca~\textrm{II} and Balmer-line activity indicators \citep[e.g.,][]{2010AJ....140.1402H, 2018MNRAS.476..908F}.
However, these existing data are mostly limited by relatively slow sampling or non-continuous temporal coverage.
The targeted cadence for this proposed survey is around 10 seconds, which is needed to capture the rapid spectral evolution during the flare phase.
Combining this fast cadence with $R\sim2000$ spectroscopy, Nautilus would be able to measure the rapidly evolving flare spectral energy distribution, the continuum-shape evolution from the near-UV/blue to the NIR, the Balmer jump, and the flux evolution of strong chromospheric features.
In terms of data quality, we aim for a signal-to-noise ratio of ${\rm SNR}=100$ per spectral pixel at $0.4~\mu{\rm m}$ per exposure/sample.

High spectrophotometric stability is also important because this program relies on measuring time-dependent relative flux changes across a broad wavelength range.
Stable relative calibration from the near-UV/blue to the NIR is needed to robustly measure the flare continuum shape, color-temperature evolution, Balmer-jump strength, and the relative energy released in different wavelength regimes.
Poorly controlled wavelength-dependent instrumental variations could mimic or obscure the chromatic spectral evolution of flares, especially for lower-energy events and for comparisons between different flare phases.
Therefore, the stable space-based observing environment of Nautilus would provide a major advantage for constructing homogeneous empirical flare templates across a large GKM stellar sample.

% Together, these measurements would connect photospheric/continuum heating, chromospheric response, and upper-atmosphere flare signatures within a single time-dependent framework, providing the key inputs needed to construct empirical flare spectral templates as a function of flare phase, flare energy, and host-star spectral type.

% \section{Data Requirements}

% % Please describe what type of data will be necessary to advance our understanding of the scientific program. Specific wavelengths, cadences, and precisions where possible.

% % We will need the fast time-resolved spectroscopic data of low-mass stars' flares.
% % The wavelength coverage we want to work with is from 0.3~$\mu m$ to 1.8~$\mu m$.
% % The targeted cadence/sampling speed of the observation is around 10-second.
% % The data quality in term of the signal-to-noise ratio (SNR) at the certain wavelength per sample of SNR=10 per sample at 0.4~$\mu m$. 

% We will need fast time-resolved spectroscopic observations of flares from low-mass stars, with wavelength coverage from $0.3~\mu{\rm m}$ to $1.8~\mu{\rm m}$. 
% The targeted cadence, or sampling speed, is around 10 seconds, which is needed to capture the rapid spectral evolution during the flare phase. 
% In terms of data quality, we aim for a signal-to-noise ratio of ${\rm SNR}=10$ per spectral pixel at $0.4~\mu{\rm m}$.

%=========================
\section{Analysis and Interpretation} 
This section describes the proposed observing strategy and the current plan for analyzing the data, with the goal of connecting the observed flare spectra to the physical properties of stellar flares on GKM dwarf stars and to their implications for planetary atmospheres around these stars.

% Please describe briefly how the data be analyzed and what would represent a successful analysis. 

\subsection{Targets selection and spectroscopic observations.}
% \begin{itemize}
% \item Pre-sample selection. 
% First, we will compile a target list of GKM dwarfs from flare-star catalogs based on Kepler/K2 and TESS observations, in order to select stars with the highest flare frequencies.  
% Currently expected numbers of sample: $\sim$100 stars/SpT. 

% \item The observing strategy we currently propose, although it may still need to be optimized, is to monitor these targets one by one for approximately 24 hours with a cadence of $\sim$10 seconds.

% \item Observations of spectroscopic standard stars and other calibration efforts will be needed to obtain well flux-calibrated data for addressing the scientific questions.
% \end{itemize}

We will first compile a target list of GKM dwarfs from flare-star catalogs based on Kepler/K2 and TESS observations to select stars with the highest flare frequencies. 
The currently expected sample size is approximately 100 stars per spectral type. 
Our currently proposed observing strategy, although it may still need to be optimized, is to monitor these targets one by one for approximately 24 hours with a cadence of $\sim$10~seconds. 

This sample size, combined with the proposed monitoring duration for each target, will provide a reasonable chance of detecting flares over a broad energy range, including relatively rare high-energy events. 
This is important because the flare frequency distributions of GKM dwarfs show that high-energy flares occur less frequently, so a large sample of flare-active stars and a sufficient observing baseline are needed to build a statistically useful data set.

Observations of spectroscopic standard stars and other calibration efforts will also be needed to obtain well flux-calibrated data for addressing the scientific questions. 
Given the expected high instrumental stability of a space-based platform such as Nautilus, we aim to construct a reference flux-calibration curve from repeated observations of spectrophotometric standard stars. 
This reference calibration can be applied to the observed flare spectra, while periodic standard-star observations will be used to monitor and correct for any time-dependent sensitivity changes over the mission lifetime.

% \subsection{Analysis of High-Cadence, Moderate- to High-Resolution Flare Spectra}
\subsection{Analysis of High-Cadence, Moderate-Resolution Flare Spectra}
\label{subsec:spectral_analysis}

We plan to investigate the spectral evolution of flares across $0.3~\mu{\rm m}$ to $1.8~\mu{\rm m}$ for flares with different energies and from stars of different spectral types. 
We will divide the flare-star sample into four spectral-type groups: G-type stars, K-type stars, early-M dwarfs (M0--M3), and mid-to-late M dwarfs (M4--M9). We separate the M dwarfs into early-M (M0--M3) and mid-to-late-M (M4--M9) groups because the transition to fully convective interiors occurs around spectral type M3--M4. 
Although this transition is not expected to be a perfectly sharp boundary, it may affect magnetic-field generation, magnetic topology, and stellar activity, and therefore provides a physically motivated division for comparing flare properties across the M~dwarf sequence.

To address the first science question in this project, we will particularly investigate how the flare energy released in the near-UV/blue wavelength range ($0.3$--$0.4~\mu{\rm m}$) is connected to the energy released at optical-to-NIR wavelengths, and how these spectral components evolve from the impulsive phase to the early-decay phase. 
The $0.3$--$0.4~\mu{\rm m}$ region contains the Balmer jump and blue continuum, which provide direct diagnostics of chromospheric heating and the depth of flare energy deposition.  This spectral region is particularly sensitive to flares that span a broad energy range, especially at low energies that may not be detected at visible and NIR wavelengths \citep[e.g.,][]{2019ApJ...871..167K}.
Moreover, we can use scaling relations to infer the flare X-ray luminosity from the near-UV emission via the Neupert effect, as discussed by \citep[][their equation 2]{2010AsBio..10..751S}, in turn, yielding an estimate of the energetic proton release in the flare \citep[][]{2005SoPh..229..135B}.  
The impact of hard protons on the chemical composition of a  planetary atmosphere can be significant, for example, leading to the significant depletion of the ozone layer \citep[][]{2010AsBio..10..751S, 2019AsBio..19...64T}.
In contrast, the optical-to-NIR continuum probes the deeper, photospheric-heating component during flares and can be used to estimate the flare color temperature and its time evolution from the impulsive phase to the early-decay phase \citep{2023ApJ...959...64H, 2025AJ....170..297L}.
By measuring the energy released in these wavelength regimes separately, we will investigate how the chromospheric and photospheric heating components are connected in addition to obtaining an estimate of the hard particle production.
We will then examine whether their relative energy budgets correlate with the total flare energy and whether these correlations vary with stellar spectral type.

% Julien:
A key analysis product will be the relative timing and energy partition among diagnostics that probe different atmospheric layers. We will measure time lags, amplitude ratios, and line-to-continuum evolution across the near-UV/blue continuum, Balmer-series emission, Ca~\textrm{II} diagnostics, H$\alpha$, and optical-to-NIR continuum. For the brightest and most active targets, line-wing variability and transient asymmetries may also provide constraints on characteristic plasma velocities and the dynamical response of the flaring atmosphere. This analysis builds on recent high-cadence UV and H$\alpha$ flare studies, and will allow Nautilus to generalize such measurements across stellar type and flare energy.

% The proposed resolving power of $R \sim 10{,}000$--$50{,}000$ will provide an additional diagnostic beyond the broadband continuum-energy budget. 
% At this resolution, we will be able to resolve and track the time evolution of key chromospheric emission lines, including the Balmer series, Paschen lines, and Ca~II H\&K lines \citep[e.g.,][]{2006PASP..118..617R, 2012ApJ...745...14S}. 

With a baseline resolving power of $R\sim2000$, Nautilus will be able to measure the time evolution of strong chromospheric emission features, including the Balmer series, Paschen lines, and Ca~II H\&K lines \citep[e.g.,][]{2006PASP..118..617R, 2012ApJ...745...14S}.
These lines are closely connected to chromospheric heating during flares and can provide complementary constraints on the density, optical depth, and dynamical response of the flaring atmosphere. 
In particular, line ratios and line-to-continuum ratios will allow us to compare the evolution of the chromospheric line-emission component with the NUV/blue and optical-to-NIR continuum components. 
The line profiles themselves can also be used to search for flare-driven broadening, asymmetries, and wavelength shifts, providing constraints on plasma motions and the time-dependent atmospheric response. 
Previous time-resolved flare spectroscopy of mid-M~dwarfs has shown that line and continuum emission do not always evolve together, making such measurements key tests for flare dynamics and radiative-hydrodynamic models \citep[e.g.,][]{2013ApJS..207...15K, 2024ApJ...961..189N}.

% Future Nautilus units with higher resolving power, $R\sim10{,}000$--$50{,}000$, would further extend this science by enabling detailed line-profile studies, including flare-driven broadening, asymmetries, and wavelength shifts that can constrain plasma motions and the time-dependent dynamical response of the flaring atmosphere.

% The line profiles themselves can also be used to search for flare-driven broadening, asymmetries, and wavelength shifts, which would provide constraints on plasma motions and the time-dependent atmospheric response during the impulsive and early-decay phases. 
% Previous time-resolved flare spectroscopy of mid-M~dwarfs has shown that chromospheric line emission and white-light continuum emission do not always evolve together, while changes in line profiles and line shifts provide important constraints on plasma motions and the time-dependent response of the flaring atmosphere \citep[e.g.,][]{2024ApJ...961..189N}. 
% These measurements also provide key observational tests for radiative-hydrodynamic flare models \citep[e.g.,][]{2013ApJS..207...15K}.
% This is especially important because previous time-resolved flare spectroscopy has shown that emission lines and continuum emission can evolve on different timescales \citep[e.g.,][]{2024ApJ...961..189N}, while changes in line profiles and line shifts can provide strong constraints on radiative-hydrodynamic flare models.

\subsection{Providing More Realistic Flare Inputs for Exoplanetary Atmospheric Evolution and Habitability Modeling}

The time-resolved flare spectra obtained in this program will be used to construct empirical flare input products for planetary atmospheric and habitability models. 
Our analysis will provide time-dependent flare spectral templates across $0.3$--$1.8~\mu{\rm m}$ as a function of flare energy, flare phase, and stellar spectral type. 
For each template, we will quantify the wavelength-dependent flare energy budget, continuum-shape evolution, flare color temperature evolution, and the relative energy released in the near-UV/blue, optical, and near-infrared regimes.

Consequently, a major product of this program will be a statistical library of flare spectral templates organized by flare energy, flare phase, and host-star spectral type. 
This library will provide a practical bridge between observed flare properties and the radiation inputs required for planetary atmospheric evolution and habitability simulations.

\subsection{Broader Applications for Stellar Physics}
Beyond the core goal of providing more realistic flare inputs for planetary atmospheric evolution and habitability modeling, this program will also enable several complementary science applications.

First, these flare spectral observations will also provide an empirical benchmark for radiative-hydrodynamic stellar flare models across the GKM sequence. 
By connecting the spectral diagnostics described in Section~\ref{subsec:spectral_analysis} with model predictions, we can test whether flare-heating properties vary systematically with stellar spectral type and magnetic activity level.
In addition, recent work suggested that the cooler-than-$9000~{\rm K}$ optical/IR flare temperatures observed on late-M dwarfs such as TRAPPIST-1 may be caused by molecular-hydrogen dissociation in the cool and dense lower atmosphere \citep{2026ApJ...998L...7S}. 
This molecular-hydrogen dissociation barrier is expected to disappear in hotter stars, possibly from early-M to K dwarfs, because their atmospheres are too warm to maintain enough molecular hydrogen. 
However, time-resolved flare-temperature measurements for K and G dwarfs are still limited, so the spectral-type transition of this effect has not been well tested observationally. 
With our broad wavelength coverage, we will measure the flare continuum temperature across different spectral types and investigate where this transition occurs.

Second, the out-of-flare and out-of-transit spectra obtained for a few hundred GKM dwarfs will provide a valuable data set for studying stellar photospheric heterogeneity. 
For each star, we can fit the quiescent spectrum with stellar atmosphere model grids, such as PHOENIX/NewEra, and compare one-component and multi-component models. 
This will allow us to test whether the observed spectrum requires heterogeneous surface components, such as cool spots, hot faculae, or magnetically active regions.
These constraints are important for interpreting exoplanet observations, because stellar heterogeneity and magnetic surface structures can affect planetary transmission and emission spectra, for example through stellar-contamination effects and changes in limb darkening \citep[e.g.,][]{2024AJ....168...82R, 2024NatAs...8..929K}.

This program would also naturally complement the Nautilus stellar heterogeneity program (Feinstein et al. in prep). 
While that program focuses on benchmark observations of quiescent and rotationally modulated photospheric heterogeneities for stellar-contamination corrections, our flare program would provide the corresponding benchmark data set for impulsive magnetic activity. 
Together, the two programs would empirically connect the photospheric, chromospheric, and upper-atmosphere behavior of GKM stars, providing the stellar inputs needed for both exoplanet atmospheric retrievals and planetary habitability modeling.
% This analysis is also relevant for exoplanet studies, because stellar heterogeneity can affect the interpretation of planetary transmission and emission spectra \citep[e.g.,][]{2024AJ....168...82R}.

% Third, the high-resolution component of the survey may enable searches for stellar coronal mass ejection (CME) candidates. 
% % With resolving power up to $R \sim 50{,}000$, we can measure transient Doppler shifts, broad wings, and blue/red asymmetries in chromospheric lines during flares \citep[][]{2019A&A...623A..49V, 2021A&A...646A..34K, 2024ApJ...961..189N, 2026NatAs..10...64N}. 
Third, this program may also enable searches for stellar coronal mass ejection (CME) candidates.
Moderate-resolution time-domain spectroscopy has already been used to search for CME-like signatures in stellar spectra; for example, \citet{2021A&A...646A..34K} used SDSS spectra with an effective resolving power of approximately $R\sim2000$ around the Balmer lines, to identify six possible CME candidates from low-mass stars through excess flux in Balmer-line wings.
With the much faster $\sim$10-second sampling proposed here, Nautilus could extend this type of search by tracking transient blue/red wing variability and broad line-wing enhancements during the rapid evolution of flares.
These measurements would help constrain the velocity structure of flare-heated or eruptive plasma and may identify candidate outflow events across GKM dwarfs.
If Nautilus units with very high-resolution spectrographs are built (for example, scanning Fabry-Perot spectrogrpahs), with a resolving power up to $R \sim 50{,}000$, this science could be further extended to detailed line-profile diagnostics, including more precise measurements of Doppler shifts, line broadening, and asymmetries in chromospheric lines during flares \citep[e.g.,][]{2019A&A...623A..49V, 2024ApJ...961..189N, 2025ApJ...978L..32L, 2026NatAs..10...64N}.
% These diagnostics will help constrain the velocity structure of flare-heated plasma and may identify candidate eruptive outflows, including possible CME-like events, across GKM dwarfs.

As a final example in this discussion, the quiescent spectra obtained in this program will provide a valuable reference data set for future studies of stellar activity cycles.
Measurements of Ca~II H\&K emission and activity indices such as the S-index will characterize the chromospheric activity level of each star at the time of observation. 
When combined with later observations or long-term monitoring programs, these measurements can serve as reference epochs for tracing activity-cycle variability in GKM dwarfs and for identifying lower-activity windows that are favorable for exoplanet transmission spectroscopy \citep[e.g.,][]{2023SSRv..219...54J,2025A&A...696A.230I, 2026arXiv260324585N}.

% As a final example in this discussion, the quiescent spectra obtained in this program will provide a valuable reference data set for future studies of stellar activity cycles.
% Measurements of Ca~II H\&K emission and activity indices such as the S-index will characterize the chromospheric activity level of each star at the time of observation. 
% These measurements will provide reference data for future activity-cycle studies of GKM dwarfs when combined with later observations or long-term monitoring programs.

In summary, our Nautilus program will provide a versatile data set that can support not only the core habitability-focused science, but also a broad range of stellar physics studies.

%=====================================================
% \section{Relevant Science Requirements}

% Please comment on science requirements to the extent it is possible. At this early stage, there is no need to demonstrate feasibility; we are mainly interested in what type of data would be important or critical for your science question. If you are not sure or do not know how to answer a question, feel free to consult with Nautilus technical team or just add "TBD" to the table.

% \begin{deluxetable}{lccl}
% \tabletypesize{\scriptsize}
% \tablecaption{Comment on necessary/optional system-level science requirements. Feel free to adjust format as it is most useful. Estimates are useful and can be refined later. \label{scireq}}
% \tablewidth{0pt}
% \tablehead{
% \colhead{Wavelength} & \colhead{Imaging} & \colhead{Spect.} & \colhead{Science Driver} 
% %\colhead{}   & \colhead{(unit)}   & \colhead{Singe/Multi}   & \colhead{}
% }
% \startdata
% 300--1800\, nm & Optional & Required & To better understand the spectral evolution from NUV-to-NIR of low-mass stars' flares. \\
% \enddata
% \tablecomments{}
% \end{deluxetable}
% \iffalse
\begin{deluxetable}{lll}
\tabletypesize{\scriptsize}
\tablecaption{Comment on necessary/optional system-level science requirements. Feel free to adjust format as it is most useful or leave entries empty if not known. Estimates are useful and can be refined later. \label{scireq}}
\tablewidth{0pt}
\tablehead{
\colhead{Requirement} & \colhead{Range} & \colhead{Science Driver}  \\
%\colhead{(unit)}   & \colhead{(unit)}   & \colhead{Singe/Multi}   & \colhead{}
}
\startdata
Photometric Filters & ... & ... \\
Spectroscopic wave. coverage & 300--1800\, nm & To better understand the spectral evolution from NUV-to-NIR of low-mass stars' flares.\\
Target Brightness [mag] & TBD &   \\
Min. Photom. Precision [ppm] & 1~ppm & White light curve photom. precision  \\
Image Res. [diff. limit] & $\sim$~0.033'' at 1.1~$\mu m$ & Derived from architecture parameters of single lens proposed by \cite{2019AJ....158...83A}. \\
Min. Sky Coverage [deg$^2$] & 2 &  Instrument design \\
Min. Contrast & 0.1 at 0.4~$\mu m$, 100 ppm at 1.5~$\mu m$ & \\
Spectral Resolving Power & $\sim2000$ & To better understand the evolution of fine spectral features during flares. \\
\hline
Monitoring Baseline [hr] & 24 & To capture unpredictable stellar flares.  \\ 
Cadence [sec] & 10 &  To obtain the well time-resolved spectral evolution of transient flare events. \\ 
Rapid Response Time [s] & TBD & \\
\hline
Data Volume & TBD  & \\
Pointing Precision [arcsec] & TBD & \\
\enddata
% \tablecomments{}
\end{deluxetable}

% Photometric Filters & ... & Explain what drives requirement\\
% Relevant Timescales [s] &  &  \\

%=========================
\section{Relevance to Nautilus and Mission Class} 

% A paragraph or two addressing why Nautilus may be the right observatory for these questions. Example justifications include: Requires very large light-collecting area; benefits from parallel observations; efficient survey execution, etc.
Nautilus is well suited to carry out the observations proposed here and to address both the core and complementary science questions. 
This program requires moderate-resolution spectroscopy with very fast cadence, at the level of $\sim$10 seconds, in order to resolve the rapid spectral evolution of stellar flares from low-mass stars. 
With its large light-collecting area, high instrumental stability, and space-based observing environment, Nautilus would be able to obtain high-quality, time-resolved flare spectra that are difficult to obtain from the ground.

\textbf{Constellation Synergy:} 
% Please comment on whether the science would benefit from implementation through a constellation. For example, would a two-stage approach (search and follow-up) with somewhat differently configured units launched in two phases be beneficial. Or would a parallelized observing capability be an advantage? 
A constellation of Nautilus units will benefit this observing program in several ways. 
If multiple units have the same instrument configuration, they can observe the same targets simultaneously, increasing the effective signal-to-noise ratio and allowing us to detect smaller flare events. 
Simultaneous observations from independent units can also reduce the risk of false positives and help distinguish real stellar variability from instrumental systematics.

% A constellation could also support parallelized wavelength coverage. 
% For example, one unit could focus on $0.3$--$1.0~\mu{\rm m}$, while another could focus on $1.0$--$1.8~\mu{\rm m}$, if this is more cost-effective or technically feasible than covering the full wavelength range with a single unit. 
% This would allow simultaneous broad-wavelength flare spectroscopy while preserving fast cadence and high data quality.

\textbf{Relevant Mission Class:} 
%Probe / Flaglet / Flagship. 
% Please comment on whether the science case may be addressed with the different class of missions, to the extent possible currently known. For reference, a Probe-class mission may be \$1B, a Flaglet may be \$1B--\$3B, and a Flagship could be \$5--\$12B.  
This science case could be partially addressed by a smaller mission class if the spectral-resolution requirement is relaxed. 
For example, a Probe-class or smaller Flaglet-class mission with $R \sim 500$--$1000$ spectroscopy over $0.3$--$1.8~\mu{\rm m}$, using a smaller lens and a less complex configuration, could still characterize the broad spectral evolution of flares across different energy scales and GKM spectral types.
Such a mission would already provide more realistic, time-dependent flare inputs for planetary atmospheric evolution and habitability modeling.

A larger Flaglet- or Flagship-class mission would be ideal for fully realizing the high-resolution components of this science case.
In particular, resolving detailed chromospheric line profiles, measuring subtle Doppler shifts and asymmetries, and detecting smaller flare events at $\sim$10-second cadence would benefit from a larger aperture, higher stability, and greater overall sensitivity.

%=========================
% \section{Relevance to NASA and Astrophysics Strategy} 

% A paragraph identifying the relevance of the science question for NASA strategic roadmaps.

% %% The "ht!" tells LaTeX to put the figure "here" first, at the "top" next
% %% and to override the normal way of calculating a float position.
% %% The asterisk after "figure" tells the compiler to span multiple columns
% %% if a two column style is selected.
% \begin{figure*}[ht!]
% \centering
% \includegraphics[width=\textwidth, height=0.5\textheight, keepaspectratio]{nautilus.png}
% \caption{Example figure.}
% \label{fig:general}
% \end{figure*}

%% Please use the acknowledgment and contribution environments. This will 
%% be anonomyized when the "anonymous" style option is used. 
\begin{acknowledgments}
We thank the Heising-Simons Foundation for supporting the Nautilus Science Case Workshop. 
\end{acknowledgments}

% \begin{contribution}

%%This section gives authors the space to recognize author contributions. The text inside this environment is NOT counted towards the total word quanta. At a minimum, manuscripts are expected to include this text:

%% But authors are expected to provide more specific details, e.g. 
%%
%%SC was responsible for writing and submitting the manuscript.
%%WWM came up with the initial research concept and edited the manuscript.
%%OTS obtained the funding and edited the manuscript.
%%EBF provided the formal analysis and validation. He also edited the manuscript.
%%GEH Supervised the undergraduates, wrote the software and administers the project github and Zenodo repositories.
%%
%% Authors can use the Contributor Role Taxonomy (CRediT) at
%% for ideas on how write a good statement tailored to their needs.

% \end{contribution}

\facilities{Nautilus \citep[][]{2019AJ....158...83A}}

% \software{}


%\appendix

\bibliography{sample701}{}
\bibliographystyle{aasjournalv7}

\end{document}